\documentclass[aps,prb,reprint,superscriptaddress,twocolumn]{revtex4-1}
\usepackage[dvipdfmx]{graphicx}
\usepackage[version=3]{mhchem}
\usepackage{braket}
\usepackage{natbib}
\usepackage{here}
\usepackage{dcolumn}
\usepackage{color}
\usepackage{bm}
\newcommand{\AffISSP}{\affiliation{Institute for Solid State Physics, University of Tokyo, Kashiwanoha, Chiba 277-8581, Japan}}
\newcommand{\AffUTPhys}{\affiliation{Department of Physics, University of Tokyo, Hongo, Tokyo 113-0033, Japan}}
\newcommand{\AffJASRI}{\affiliation{Japan Synchrotron Radiation Research Institute, 1-1-1 Kouto, Sayo, Hyogo 679-5198, Japan}}
\newcommand{\AffIMR}{\affiliation{Institute for Materials Research, Tohoku University, Sendai, Miyagi 980-8577, Japan}}
\newcommand{\AffRIKEN}{\affiliation{RIKEN SPring-8 Center, Sayo-cho, Hyogo 679-5148, Japan}}
\newcommand{\AffCSRN}{\affiliation{Center for Spintronics Research Network, Tohoku University, Sendai, Miyagi 980-8577, Japan}}
\newcommand{\AffIMS}{\affiliation{Institute for Molecular Science, Okazaki, Aichi 444-8585, Japan}}
\usepackage{amsmath}	
\begin{document}
\title{Element-selective tracking ultrafast demagnetization process in Co/Pt multilayer thin films by the resonant magneto-optical Kerr effect}
\author{Kohei~Yamamoto}
\email{yamako@ims.ac.jp}
\homepage{http://sites.google.com/site/yamakolux/}
\AffISSP\AffUTPhys\AffIMS
\author{Souliman El Moussaoui}\AffISSP
\author{Yasuyuki~Hirata}\AffISSP\AffUTPhys
\author{Susumu~Yamamoto}\AffISSP\AffUTPhys
\author{Yuya~Kubota}\AffJASRI\AffRIKEN
\author{Shigeki~Owada}\AffJASRI\AffRIKEN
\author{Makina~Yabashi}\AffRIKEN\AffJASRI
\author{Takeshi~Seki}\AffIMR\AffCSRN
\author{Koki~Takanashi}\AffIMR\AffCSRN
\author{Iwao~Matsuda}\AffISSP\AffUTPhys
\author{Hiroki~Wadati}
\altaffiliation{Present address: Department of Material Science, University of Hyogo, Kamigori-cho, Hyogo 678-1297, Japan}
\AffISSP\AffUTPhys
\date{\today}
\begin{abstract}
We examined the photo-induced dynamics of ferromagnetic Co/Pt thin films demonstrating perpendicular magnetic anisotropy with element specificity using resonant polar magneto-optical Kerr effect measurements at Pt~N${}_{6,7}$ and Co~M${}_{2,3}$ edges with an x-ray free electron laser.
The obtained results showed a clear element dependence of photo-induced demagnetization time scales:\ $\tau_\textrm{demag.}^\textrm{Co}=80\pm60~\textrm{fs}$ and $\tau_\textrm{demag.}^\textrm{Pt}=640\pm140~\textrm{fs}$.
This dependence is explained by the induced moment of the Pt atom by current flow from the Co layer through the interfaces.
The observed magnetization dynamics of Co and Pt can be attributed to the characteristics of photo-induced Co/Pt thin film phenomena including all-optical switching.
\end{abstract}
\maketitle

The dynamics of photo-induced magnetization has been a central issue in the physics of non-equilibrium spin states and also in the development of ultrafast information devices~\cite{Kirilyuk2010}. All optical switching (AOS) of magnetization has been one of the key phenomena, and was first observed in ferrimagnetic thin films such as GdFeCo\cite{Stanciu2007,Mangin2014}, followed by recent discoveries of ferromagnetic granular films of FePt and ferromagnetic Co/Pt multilayers~\cite{Lambert2014}. 
These AOS materials are magnetized in the out-of-plane direction~\cite{Kimel2014,Lambert2014}.
It has now become possible to synthesize a variety of AOS materials depending on their elements and structure. 
Despite extensive experimental observation, determination of the underlying physics has been elusive due to the complexity of photo-induced magnetic phenomena. 
AOS materials are inherently composed of more than one magnetic elements and transient states at the individual sites are determined by various internal degrees of freedoms , such as angular momenta, spin densities of states, and lattice vibrations. 
It has become essential in experiments to disentangle ultrafast spin dynamics at each magnetic element.
 
Element-selective investigation of magnetic materials have conventionally been made using x-ray magnetic circular dichroism (XMCD) that probes absorption edges of the target elements~\cite{Stohr2006}. 
The ultrafast time-resolved measurement of XMCD can be made with a free electron laser (FEL) or a laser-slicing synchrotron radiation source in the soft x-ray region, i.e., $h\nu\sim700-2000~\textrm{eV}$, particularly for L${}_{2,3}$ edges of 3d elements and M${}_{4,5}$ edges of 4f elements~\cite{Malvestuto2018,Lopez-Flores2013,Bergeard2014,Eschenlohr2014,Radu2015,Radu2011,Takubo2017,Yamamoto2015a,Boeglin2010}.
These experiments helped to improve our understanding of photo-induced magnetization dynamics for ferrimagnetic AOS materials such as GdFeCo~\cite{Mentink2012,Radu2015,Radu2011}.
Note that M${}_{4,5}$ edge of Gd ($\sim 1400~\textrm{eV}$) and L${}_{2,3}$ edge of Fe ($\sim 700~\textrm{eV}$) can be probed in a soft x-ray beamline.
However, due to the limitations of photon energy, some key elements, such as 4d and 5d elements, cannot be investigated by soft x-rays - thus experiments with differing techniques are sometimes required for individual elements, raising the potential for experimental uncertainty between measurements. 
This is the case for ferromagnetic AOS materials and 3d and 5d alloy materials~\cite{KY1810.02551}.

The absorption edges of shallow core levels in the extreme ultraviolet (EUV) region, i.e., $h\nu\lesssim$ 150~eV, cover a wide range of elements.
A time-resolved measurement with an ultrashort pulsed EUV source can be applied to elements including 5d elements. This enables the direct comparison of the magnetization dynamics of various elements ~\cite{Willems2015}. 

The magneto-optical Kerr effect (MOKE) is detected as the rotated plane of polarization of the reflected beam corresponding to the sample magnetization.
X-ray MOKE (XMOKE) can be enhanced at the absorption edges and get element selectivity as in the case of XMCD. 
XMOKE of 3d elements in EUV reflect 3p spin-orbit splitting and 3d spin polarization and can determine exchange-splitting and spin-wave excitation~\cite{Erskine1975}.
From a technical standpoint, a linearly polarized x-ray beam originating from a linear undulator is available for probing XMOKE without using a specific undulator that generates a circular polarized x-rays.
In addition, EUV is surface sensitive compared to soft x-rays and provides complementary information.
Transverse XMOKE in EUV range was adopted for magnetization measurements using high harmonic generation (HHG). This can be performed without polarization analysis, and only in-plane magnetization dynamics was observed~\cite{Mathias2012,La-O-Vorakiat2009,Hofherr2018}.

In this letter, we report on the spin dynamics of ultrafast demagnetization phenomenon in perpendicularly magnetized AOS Co/Pt multilayer material by performing time-resolved XMOKE (trXMOKE) measurements using an EUV FEL source~\cite{Yamamoto2015a}. 
We employ polar XMOKE setup for observing out-of-plane magnetization.
The absorption spectra show the apparent absorption peak at the Co M edge and Fano-resonance at the Pt N edge, both of which exist in the EUV region. 
By optimizing photon energy with the energy-tunable FEL source, spin dynamics at the Co and Pt sites were traced element-selectively. 
Femtosecond demagnetization is initiated at the Co sites and then at the Pt sites. 
These results indicate spin transports from the Co to Pt layer during the ultrafast process.


\begin{figure}
\begin{center}
  \includegraphics[clip,width=8cm]{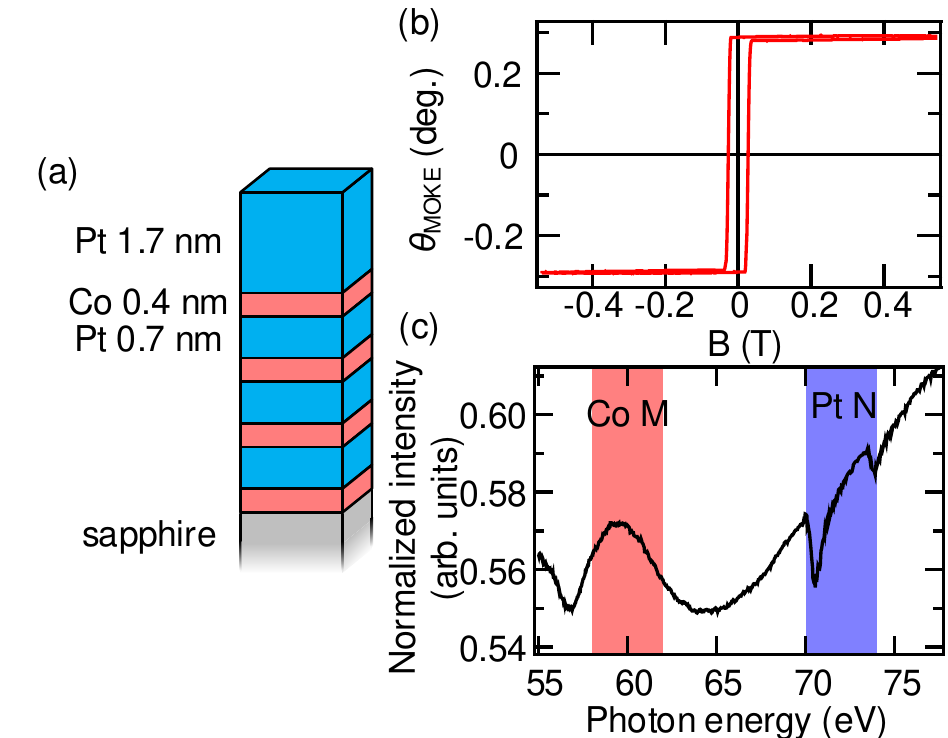}
  \caption{Sample description and characterization. (a) The multilayer structure of Pt(1.7~nm)/[Co(0.4~nm)/Pt(0.7~nm)]${}_3$/Co(0.4~nm) on a Sapphire($11\overline{2}0$) substrate. (b) Magnetic hysteresis loop measured by polar MOKE. (c) X-ray absorption spectrum of Co M${}_{2,3}$ and Pt N${}_{6,7}$ edges.}
\label{fig:XAS}
  \end{center}
  \vspace*{-0.5cm}
\end{figure}

We used a Pt(1.7~nm)/[Co(0.4~nm)/Pt(0.7~nm)]${}_3$/ Co(0.4~nm) multilayer thin film synthesized on Sapphire($11\overline{2}0$) via the sputtering method.
This multilayer structure is depicted in Fig.~\ref{fig:XAS}(a).
This sample exhibited perpendicular magnetic anisotropy.
We measured the magnetic hysteresis loop by visible light polar MOKE with a wavelength of 680 nm and the result indicates the out-of-plane ferromagnetic component as shown in Fig.~\ref{fig:XAS}(b).
Figure~\ref{fig:XAS}(c) shows the x-ray absorption spectrum of our sample, which was obtained by the total electron yield method at the synchrotron radiation facility UVSOR BL5B using horizontally polarized x-rays.
The absorption spectrum with Co~M${}_{2,3}$ (60~eV) and Pt~N${}_{6,7}$ (72~eV) edges is similar to that previously reported for CoPt~\cite{Willems2015,Shishidou1997}, including the dip of Pt~N edge resulting from the Fano effect.

\begin{figure}
\begin{center}
  \includegraphics[clip,width=8cm]{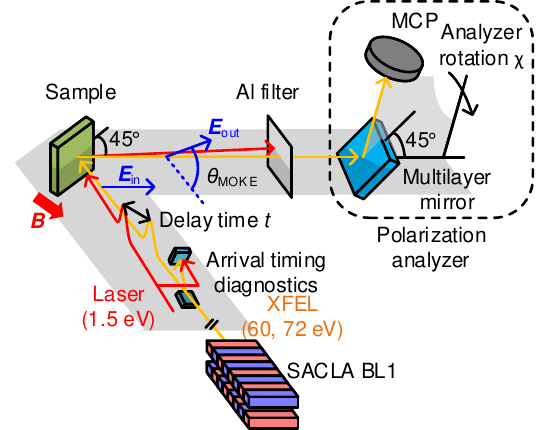}
  \caption{Schematic of the trXMOKE setup built at SACLA BL1.
The EUV and pump laser pulses were incident on the sample.
The time difference between these two beams was measured using arrival timing diagnostics.
Al filter cut the reflected visible pump laser.
The polarization of reflected x-ray was analyzed by the rotating polarization analyzer.
}
\label{fig:setup}
  \end{center}
  \vspace*{-0.5cm}
\end{figure}

\begin{figure*}
\begin{center}
  \includegraphics[clip,width=17cm]{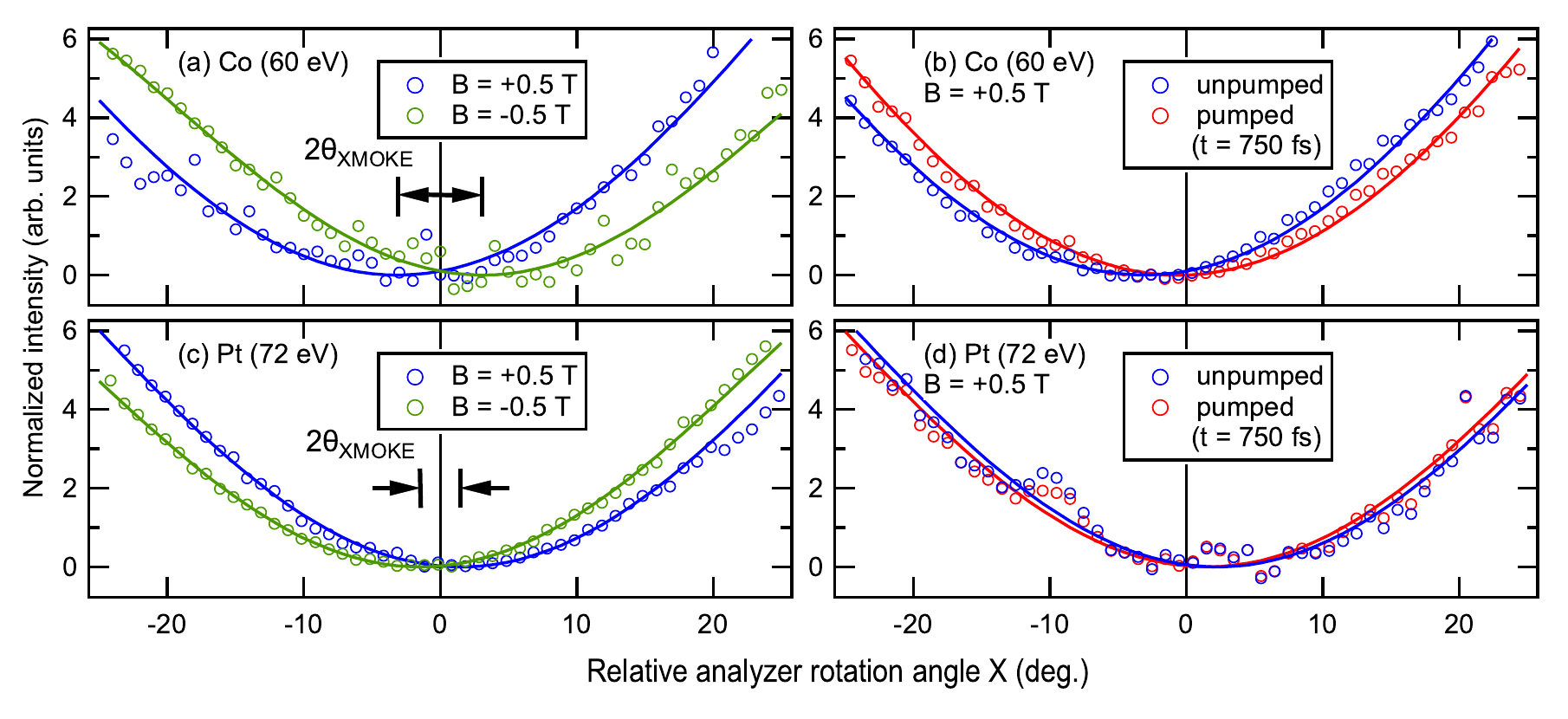}
  \caption{Reflectivity intensity are plotted as functions of polarization analyzer angle $\chi$ for Co~M${}_{2,3}$ (60~eV) (a) and Pt~N${}_{6,7}$ (72~eV) (c) edges. The pump effect was observed as the shift of the curves at the both Co~M (b) and Pt~N (d) edges.}
\label{fig:curves}
  \end{center}
  \vspace*{-0.5cm}
\end{figure*}

We used an EUV FEL ($<$ 150 eV) free electron laser at BL1~\cite{Owada2018a} of SACLA~\cite{Ishikawa2012}, an FEL facility in Japan. 
The schematic drawing of the experimental setup is shown in Fig.~\ref{fig:setup}.
TrXMOKE measurements were performed using the pump-probe technique~\cite{Yamamoto2018b} to capture the magnetic moment dynamics with element specificity~\cite{Yamamoto2014}.
FEL pulses were incident at an angle of 45~deg.
The polarization of incident FEL pulses was linear and the polarization plane of the reflected FEL was analyzed by rotating analyzer-ellipsometry~\cite{Kubota2017a,Yamamoto2018b}.
The polarization analyzer consisted of a multilayer mirror and a micro channel plate (MCP), both of which rotate simultaneously.
The Mo/Si multilayer mirror reflects scattered FEL pulses from the sample with an incident angle of 45 deg., with the reflected FEL pulses hitting the MCP.
Ellipsometry curves as a function of rotation angle $\chi$ can be obtained by rotating the ellipsometry analyzer.
The details of the method used in analyzing polarization are described in ref.~\cite{Kubota2017a,Yamamoto2018b}.
The photon energy of FEL was 60~eV for the Co~M${}_{2,3}$ edge and 72~eV for the Pt~N${}_{6,7}$ edge, which was determined from our x-ray absorption measurement.
The wavelength, fluence, and spot size of the optical laser were 1.5~eV (800~nm), 10.6~$\textrm{mJ/cm}^2$, and $200~\mu\textrm{m}\times 180~\mu\textrm{m}$, respectively.
A magnetic field of 0.5~T was applied throughout the experiment and the magnetization of the sample was saturated.
Both the FEL and optical laser were incident nearly coaxially and reflected optical laser light was cut by the aluminum filter in front of the polarization analyzer.
The repetition rate or frequency of FEL and optical laser pulses were 60 Hz and 30 Hz. We observed pumped and unpumped states by using every second pulse, which enabled detection of the pump effect precisely. 
We confirmed that the sample was not damaged by observing the unpumped states.
An arrival timing diagnostic system between the FEL and optical laser pulses was introduced~\cite{Owada2018} to compensate for the arrival timing jitter.
The total time resolution of $\sim 70$~fs was achieved in our measurement.


Figures~\ref{fig:curves} show MCP intensity as a function of rotation angle $\chi$.
The ellipsometry curves were analyzed by fitting with the following equation:
\begin{equation}
 I(\chi)=A_0+A_1\sin(\alpha(\chi-\chi_0)+3\pi/2).
\end{equation}
The result obtained under opposite magnetic fields is shown in Figs.~\ref{fig:curves}(a, c).
$\theta_\textrm{XMOKE}$ can be determined from the difference between these curves: $2\theta_\textrm{XMOKE}=\left|\chi_0(+B)-\chi_0(-B)\right|$.
$\theta_\textrm{XMOKE}$ of Co and Pt were determined as $\theta_\textrm{XMOKE}^\textrm{Co}=3.1 \pm 0.3^\circ$ and $\theta_\textrm{XMOKE}^\textrm{Pt}=1.5 \pm 0.3^\circ$.
$\theta_\textrm{XMOKE}$ of the absorption edges are 5 to 10 times larger than $\theta_\textrm{MOKE}$ of 680~nm as shown in Fig.~\ref{fig:XAS}(b).
The observed enhancement of $\theta_\textrm{XMOKE}$ at the absorption edges clearly shows the resonant effect, which is similar to the previous XMOKE results \cite{Yamamoto2014}.

Figures~\ref{fig:curves}(b, d) demonstrate the ellipsometry curves for the pumped and unpumped samples in typical cases of trXMOKE.
We determined $\theta_\textrm{XMOKE}$ with the relationship $\theta_\textrm{XMOKE}(t)=|\chi_0(+B)-\chi_0(-B)|/2-|\chi_0^\textrm{unpumped}-\chi_0^\textrm{pumped}(t)|$, where $\chi_0(+B)$ and $\chi_0(-B)$ are values of the static measurements.
We detected pump effect through the shift of ellipsometry curves and evaluated this shift by changing delay time $t$.

\begin{figure}
\begin{center}
  \includegraphics[clip,width=8cm]{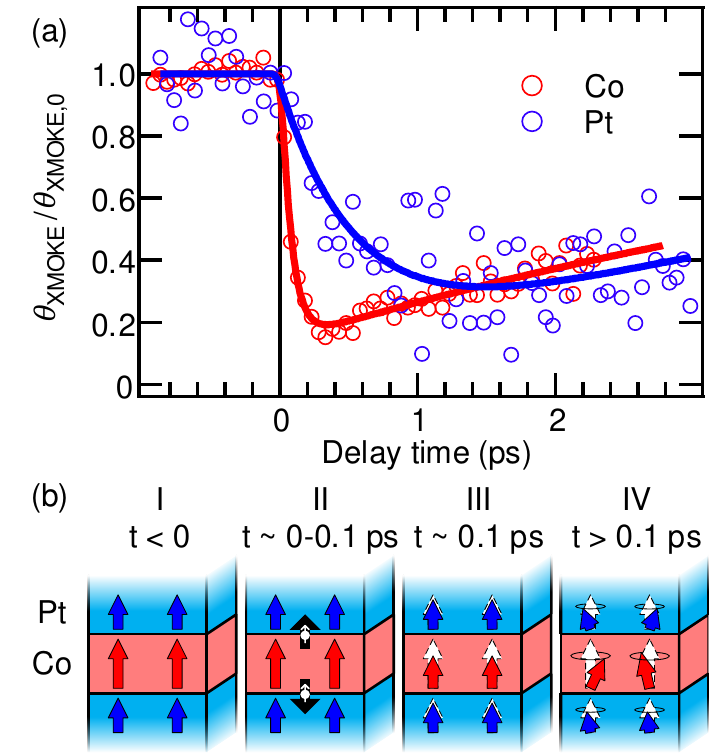}
  \caption{Photo-induced dynamics of Co and Pt magnetization dynamics obtained by trXMOKE. (a) Kerr rotation angle at Co and Pt.
Kerr rotation angle $\theta_\textrm{XMOKE}$ is normalized to the initial value $\theta_\textrm{XMOKE,0}$.
Solid lines indicate the results of fitting.
(b) The schematics of unpumped and pumped states. I: initial states. II: flow of superdiffusive current. III: fast reduction of Co moment. IV: incoherent transverse spin excitation with a high spin temperature.}
\label{fig:trXMOKE}
  \end{center}
  \vspace*{-0.5cm}
\end{figure}

Figure~\ref{fig:trXMOKE}(a) exhibits the result of trXMOKE measurement changing the delay time systematically.
The results are normalized by the Kerr rotation angle before excitation, $\theta_\textrm{XMOKE,0}$.
A clear element-dependence of the photo-induced demagnetization process can be seen.
We evaluated time scales of Co and Pt with an exponential function 
\begin{equation*}\begin{split}
\theta_\textrm{XMOKE}&=1+A\left[1-\exp(-t/\tau_\textrm{demag.})\right]\\
&\quad\left[a+(1-a)\exp(-t/\tau_\textrm{recovery})\right]H(t),
\end{split}\end{equation*}
where $H(t)$ is the Heaviside step function.
The time scales were determined as $\tau_\textrm{demag.}^\textrm{Co}=80\pm60~\textrm{fs}$ and $\tau_\textrm{demag.}^\textrm{Pt}=640\pm140~\textrm{fs}$.

The results of this work can be explained in the similar scheme of trXMCD of FePt~\cite{KY1810.02551}.
Prior to laser excitation, Co and Pt magnetic moments are coupled ferromagnetically and the sample is perpendicularly magnetized as shown in step I of Fig.~\ref{fig:trXMOKE}(b).
A pump laser with photon energy of 1.5~eV excites Co and Pt electrons.
Pt~d electrons lie in deeper energy level from $E_F$ and the difference in spin polarization of ferromagnetic Co and paramagnetic Pt around $E_F$ would enhance the excitation of Co majority spin.
The excited majority spin of Co hot electrons transport to the Pt site via superdiffusive transport~\cite{Battiato2012,Schellekens2014} or optical intersite electron transfer~\cite{Muller2009a} at the interfaces.
Majority spin of Co can be excited into a mobile sp-like band~\cite{Willems2015} as in the case of Ni~\cite{Battiato2010} and Pt has relatively localized feature of d electron. This may cause asymmetric behavior of electron diffusion.
Transport of majority spin from Co to Pt layer leads to faster and slower reduction of Co and Pt magnetization, respectively, as depicted schematically in steps II and III of Fig.~\ref{fig:trXMOKE}(b).
Energy and angular momentum dissipate to other freedom  of lattice and spin excitation~\cite{Carva2017}. Finally, incoherent transversal spin excitation appears with high spin temperature as shown as step IV.

Our results demonstrate long-lived Pt momentum and the existence of the modulation of Pt and Co momenta ratio in Co/Pt.
AOS has not been observed in single-element materials but has been detected in alloys. This suggests the importance of exchange coupling and spin-orbit interaction at interfaces~\cite{Hellman2017}.
There is a possibility that the observed modulated magnetic states of sublattices are the origin of the mechanism of AOS or other photo-induced phenomena.

Compared to previous reports~\cite{Willems2015,Hofherr2018}, our observed $\tau_\textrm{Pt}$ is relatively slow.
We note that there are some differences from previous reports and our measurement.
The first point is the magnetic character of the samples.
According to the XMCD study of FePt with various order degrees, the magnetic feature of thin films changes from isotropic to anisotropic when its order degree is higher, via strong spin-orbit coupling of Pt~\cite{Ikeda2017}.
In the previous reports of trXMCD~\cite{Willems2015} and transverse trXMOKE~\cite{Hofherr2018} of 3\textit{d} and Pt~O edges, the studied FePt and Co/Pt samples were magnetized along the in-plane direction.
This indicates that the samples have more isotropic magnetic character than our sample.
The change of anisotropy may be related to lattice-spin coupling and thus contribute to our relatively slow dynamics of Pt in addition to the electron transfer process from Fe to Pt layer.
The second point is the selection of absorption edges.
In the 50-75 eV range, there are N${}_{6,7}$ and O${}_3$ edges of Pt and M${}_{2,3}$ edges of Fe and Co.
The small energy difference of absorption edges can cause mixing of information of each other elements~\cite{Antonov2001} and the degree of mixing can vary depending on the position of absorption edges.
We used Pt N${}_{6,7}$ edges, which is most isolated from Co M edges and exhibit the largest XMCD signals in the EUV range.
This was aimed at better element-selectivity for Pt magnetic moments.
The third point is the XMCD intensity of the absorption edges.
According to the spectra reported~\cite{Willems2015,Honolka2009,Shishidou1997}, Pt N${}_{6,7}$ edge XMCD intensity is larger than Pt~O edges.
By using larger XMCD at Pt N${}_{6,7}$, it is expected information of 3d metal and Pt will be more clearly separated.

In conclusion, we succeeded in conducting element-specific observation of photo-induced magnetization dynamics of a perpendicularly magnetized Co/Pt multilayer thin film.
The demagnetization time scales of Pt and Co were found to be $\tau_\textrm{demag.}^\textrm{Co}=80\pm60~\textrm{fs}$ and $\tau_\textrm{demag.}^\textrm{Pt}=640\pm140~\textrm{fs}$.
We suggest that these observed dynamics can be realized by majority spin Co electron transport into the Pt layer.
The different photo-induced dynamics of Co and Pt lead to the modulation of the ratio of Co and Pt magnetization.
This transient state can affect photo-induced dynamics through interfacial exchange coupling or electron transportation in the photo-induced process including AOS.
Finally, we emphasize that ultrafast dynamics of Pt and Co moments have been determined simply with a single instrument using the EUV-FEL-based trXMOKE. This new approach will elucidate the ultrafast magnetization dynamics of a vast variety of advanced complex materials.

\begin{acknowledgments}
This work was partially supported by Toray Science and Technology Grant of Toray Research Foundation and Grant-in-Aid for Young Scientists (B) (No. 17K14334) of the Japan Society for the Promotion of Science (JSPS).
 The FEL experiments were performed at SACLA with the approval of the Japan Synchrotron Radiation Research Institute (JASRI) (Proposal No. 2018B8040, 2018B8052).
  K.~Y. acknowledges the support from ALPS program of the University of Tokyo.
\end{acknowledgments}

%

\end{document}